\documentclass{webofc}

\usepackage[varg]{txfonts}   
\usepackage{hyperref}
\usepackage{url}
\usepackage{wrapfig}
\hypersetup{colorlinks=true,citecolor=blue,urlcolor=blue,linkcolor=blue}
\newcommand{\eps}{\epsilon}
\def\bea {\begin{eqnarray}}
\def\eea {\end{eqnarray}}
\begin{document}
\title{Hadron resonance gas is not a good model for hadronic matter in a strong magnetic field.}
%
%

\author{\firstname{Pasi} \lastname{Huovinen}\inst{1}\fnsep\thanks{\email{pasi.huovinen@uwr.edu.pl}} \and
        \firstname{Michał} \lastname{Marczenko}\inst{1} \and
        \firstname{Michał} \lastname{Szymański}\inst{2} \and
        \firstname{Bithika} \lastname{Karmakar}\inst{1} \and
        \firstname{Pok Man} \lastname{Lo}\inst{2} \and
        \firstname{Chihiro} \lastname{Sasaki}\inst{2,3} \and
        \firstname{Krzysztof} \lastname{Redlich}\inst{2,4}
}

\institute{Incubator of Scientific Excellence---Centre for Simulations of
  Superdense Fluids, University of Wroc\l{}aw, 50204 Wroc\l{}aw, Poland
\and
           Institute of Theoretical Physics, University of Wroc\l{}aw,
           50204 Wroc\l{}aw, Poland 
\and
           International Institute for Sustainability with Knotted Chiral Meta
           Matter (WPI-SKCM$^2$), Hiroshima University, Higashi-Hiroshima,
           Hiroshima 739-8526, Japan
\and
           Polish Academy of Sciences PAN, 50449 Wroc\l{}aw, Poland
          }

\abstract{We study the effect of magnetic field on particle yields and charge
fluctuations in hadron resonance gas. We argue that the big changes in
the proton yield and baryon number susceptibility are due to ill-defined
description of higher-spin states, and that because of detailed balance,
neutral resonances must be affected by the field too.
}
\maketitle
\section{Introduction}
  \label{intro}

The behaviour of QCD matter in strong magnetic field is of both
practical and theoretical interest. Practical, since the magnetic
fields generated in non-central heavy-ion collisions at
ultrarelativistic energies are among the strongest in the
universe~\cite{Kharzeev:2007jp}, and theoretical, since it can be
calculated from first principles using lattice QCD (LQCD) methods.
Recently, the LQCD calculations of the fluctuations of conserved
charges in finite magnetic field have been of particular interest (see
e.g.~Ref.~\cite{Ding:2025jfz}). It has been observed that a finite
magnetic field has a nontrivial effect on those observables.

The hadron resonance gas (HRG) model has been successful in describing
the LQCD results on the EoS and fluctuations of conserved charges at
vanishing magnetic field~\cite{Bollweg:2021vqf}.
In~\cite{Endrodi:2013cs}, the HRG model was generalised to include the
effects of finite magnetic field. In this contribution we calculate
the fluctuations of conserved charges and particle yields in a finite
magnetic field using the HRG model~\cite{Marczenko:2024kko}. We argue
that the strong dependence on the field strength is an artefact of the
model assumptions, and appears in a region where the model contradicts
its own premises.

\section{Hadron resonance gas in finite magnetic field}

The HRG model is based on the approximation
that in interacting hadron gas interactions mediated by resonances
dominate. If these resonances are narrow, their contribution to the
equation of state of the gas can be well approximated by describing
them as noninteracting particles with the pole mass of the
resonance. Thus the interacting system can be approximated as a gas of
free hadrons and resonances. Furthermore, the hadrons and resonances
are assumed to be pointlike and structureless, and the usual kinetic
theory definitions of thermodynamic quantities apply.

In the presence of a constant external magnetic field $eB$ pointing
along the $z$ direction, the system undergoes Landau quantisation in
the $xy$ plane~\cite{Andersen:2014xxa}. Consequently, the dispersion
relation for a charged particle $(Q \neq 0)$ becomes
\begin{equation}
\label{energy}
    \eps = \sqrt{p_z^2 + m^2 + 2|Q|B\left(l + 1/2 - s_z\right)}\textrm,
\end{equation}
where $s_z$ is the $z$-component of the particle's spin and
$l \in \{0,1,2,\ldots \}$ numbers the Landau levels. We note that the
dispersion relation in Eq.~\eqref{energy} is exact for structureless
spin-$0$ and spin-$1/2$ particles~\cite{Andersen:2014xxa}, but a
similar relation can be derived for spin-$1$ and $3/2$
particles. Likewise, the integration over transverse momenta in the
thermodynamic integrals is replaced by a summation over Landau levels.

As seen in Eq.(\ref{energy}), even the lowest Landau level gives a
positive contribution to the energy of pions (and other spin-0
particles). Thus the pion density in fixed temperature decreases with
increasing magnetic field. The contribution of the lowest Landau
level to the energy of spin-$1/2$ particles is zero, but we observe a
small increase in their densities with increasing magnetic field due
to quantisation of the energy levels (summation instead of
integration), whereas the contribution to the energy of spin-$1$ and
$3/2$ particles is negative, and we observe a strong increase of the
densities of $\rho$ and $\Delta$ with increasing magnetic
field~\cite{Marczenko:2024kko}.

\begin{figure*}[b]
\centering
\includegraphics[width=0.46\textwidth,clip]{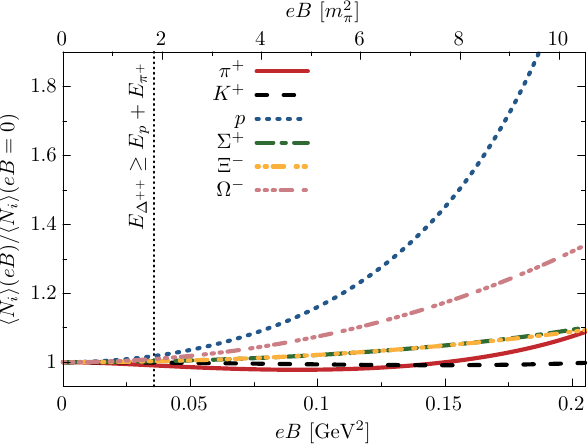}
  \hfill
\includegraphics[width=0.46\textwidth,clip]{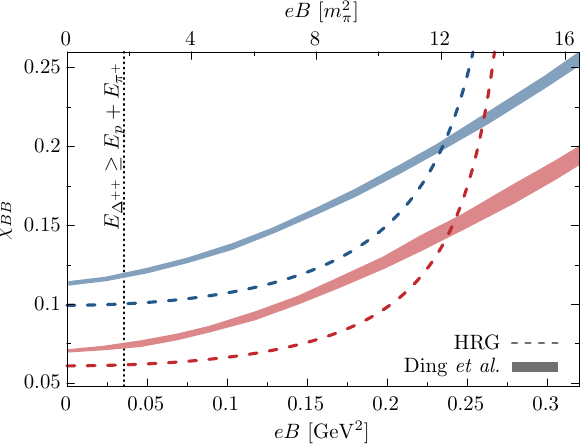}
\caption{Particle yields normalised to the yields at vanishing
  magnetic field at $T=155~$MeV (Left) and net-baryon number
  susceptibility at $T=0.145$ (red) and $0.155~$GeV (blue) (Right). \vspace*{-5mm} }
\label{fig-1}
\end{figure*}

Particle densities are not observables in heavy-ion collisions since
final observed particle yields contain contributions from all decayed
resonances. 
In the left panel of Fig.~\ref{fig-1}, we show the particle
yields after decays, normalised to the yields at vanishing magnetic
field. Even if the density of thermal pions decreases, they receive a
large contribution from rhos, and thus the dependence of the
pion yield on magnetic field is weak. The kaons behave similarly, but
protons get a large contribution from Deltas, and therefore the
proton yield strongly depends on the magnetic field. The other
spin-$1/2$ baryons, $\Sigma$ and $\Xi$, have way weaker dependence since
there are way fewer known strange than non-strange resonances.

In a thermal medium, the second-order fluctuations and correlations of
conserved charges are quantified by the generalised susceptibilities.
In the right panel of Fig.~\ref{fig-1} we show the HRG model
net-baryon number susceptibility $\chi_{BB}$ as a function of the
strength of the magnetic field, and compare it to the LQCD
result~\cite{Ding:2025jfz} at T=145 and 155 MeV. We find that
the HRG model systematically underestimates the LQCD result for
$eB \leq 0.22~\rm GeV^2$ at both temperatures. If all the hadron
states predicted by relativistic quark model are included in the HRG
model, the lattice $\chi_{BB}$ is reproduced, but independent of the
particle list, HRG model overshoots the lattice result at large values
of $eB$. This behaviour is again due to the $s\geq 3/2$ resonances,
which contribution ranges from $75\%$ at vanishing $eB$ to $95\%$ at
$eB=0.3~\rm GeV^2$.

At this point it is worth remembering that Eq.~\eqref{energy} neglects
the compositeness of the states. Issues can arise when the scale of
magnetic field resolves the structure of hadron states, i.e.,
$eB > m_\pi^2$. For example the contribution of the lowest Landau
level to the energy of high-spin particles is negative, see
Eq.~(\ref{energy}). Therefore in strong enough magnetic field the
dispersion relation of, say, $\Delta^{++}$, becomes complex signalling
instability. Likewise the concept of a resonance can become
questionable: For the resonance to be able to decay, its energy must
be larger than its daughter particles. E.g. for $\Delta^{++}$,
$E_{\Delta^{++}} \geq E_p + E_{\pi^+}$. But since the lowest energy
level of $\Delta^{++}$ ($\pi^+$) decreases (increases) with increasing
strength of the magnetic field, this requirement is no longer valid
once $eB \gtrsim 0.0356$ GeV$^2$! Thus in stronger field it is not clear
what is a resonance and what is a ground state hadron. In
Fig.~\ref{fig-1} this limit is shown as a vertical dotted line. Below
this line, the effect of the magnetic field is negligible, but above
it, our results are not to be trusted.

\section{Detailed Balance}
  \label{det_balance}

\begin{wrapfigure}{r}{0.5\textwidth}
  \begin{center} \vspace*{-11.5mm}
    \includegraphics[width=0.5\textwidth,clip]{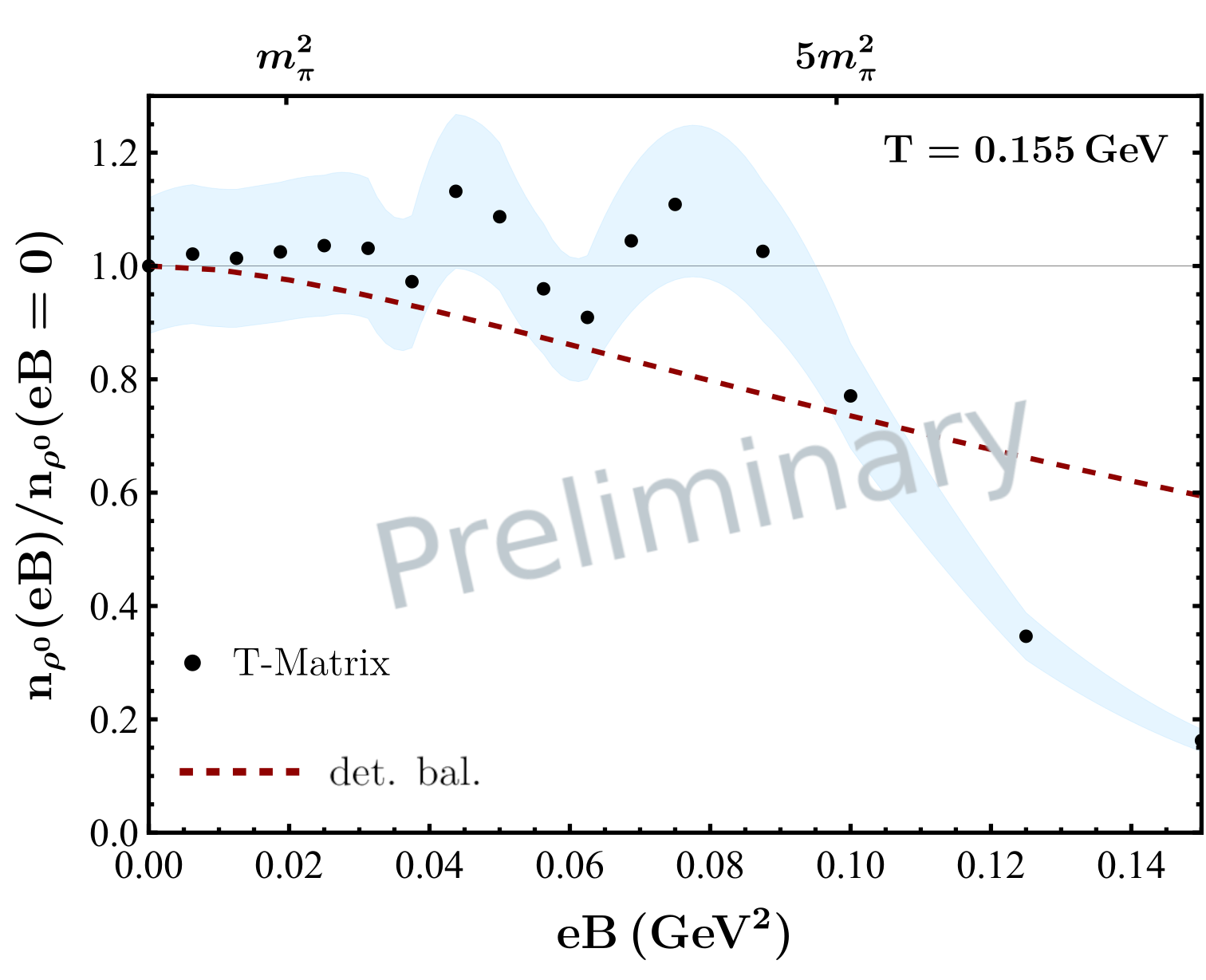}
  \end{center}
  \caption{$\rho^0$ meson density at $T = 155$ MeV using the detailed
    balance and T-matrix approaches normalised to densities at $eB=0$.
 \vspace*{-5mm}}
\label{fig-2}
\end{wrapfigure}
  
We assume neutral particles to have zero magnetic moments. Therefore,
they are not affected by the magnetic field at all. However, in HRG in
equilibrium detailed balance prevails, e.g.,~the decay rate of, say,
rho mesons is equal to the scattering rate of pions forming them. The
scattering rate depends on the densities and, as mentioned previously,
pion density decreases with increasing magnetic field. On the other hand,
decay rate depends on the resonance width, and it is known that a weak
magnetic field causes a negligibly small change in the width of
$\rho^0$~\cite{Bandyopadhyay:2016cpf}. Thus in the presence of a
magnetic field, the $\rho^0$ decay rate changes only if the $\rho^0$
density changes, and the change in rate is proportional to the change
in density.

Unfortunately, we cannot evaluate the scattering rate of pions using
the conventional kinetic theory calculation, since particles in
magnetic field do not have well-defined momenta. Instead, we make a
bold assumption that the magnetic field affects the pion scattering
rate (and thus the $\rho^0$ production rate) only by changing their
densities. We can evaluate an effective pion chemical potential such
that $n_\pi(T,\mu_\pi,B=0) = n_\pi(T,B)$. With the above-mentioned
considerations, detailed balance requires that $\rho^0$ obtains a
chemical potential, which is a sum of pion chemical potentials:
$\mu_{\rho^0} = \mu_{\pi^+} + \mu_{\pi^-} = 2\mu_{\pi^+}$. Similar 
arguments apply to $\Delta^0$~\cite{Marczenko:2024kko}.

The $\rho^0$ density corresponding to this chemical potential, and
scaled with its equilibrium density, is shown in in
Fig.~\ref{fig-2}. Unlike the expectation that neutral particles should
not be affected by the magnetic field, $\rho^0$ density decreases with
increasing magnetic field, the effect being the stronger the colder
the system. Since the densities of neutral resonances are affected via
this mechanism, it is questionable what the proper treatment of
charged resonances is. Similar detailed balance approach leads to very
different sensitivity to the magnetic field than the treatment of
resonances as elementary particles according to Eq.~(\ref{energy}).

\section{Hadronic structure and thermal yields}

The dispersion relation in Eq.~\eqref{energy} assumes structureless, i.e. pointlike, particles, and thus accounts only for very narrow resonances. To incorporate more general particle decay dynamics, which are essential for a reliable description of hadron yields at higher magnetic fields, one should employ a consistent model of hadronic interactions.
As a first step towards such a model, we consider a $\pi-\rho$ gas where
the $\rho_0 \pi^+ \pi^-$ interaction is described by the Lagrangian
$ \mathcal L_{int} = -g \,\partial_\mu \vec{\rho_\nu} \cdot \partial^\mu \vec {\pi}\times \partial^\nu \vec{\pi}, $
with $g=20.72$ GeV${^{-2}}$, which is fixed from the vacuum decay width
$\Gamma_{\rho \rightarrow \pi \pi}$= 150 MeV ~\cite{Ghosh:2017rjo}.
The one loop self energy ($\Pi^{\mu\nu}$) of the $\rho_0$ meson is
evaluated in presence of an arbitrary magnetic field, employing the
Schwinger propagator for the loop pions. The $\rho_0$ density in an
external magnetic field is then obtained by constructing the in-medium spectral function.
$\text{Re}~ \Pi$ has not been taken into account in the spectral
function. Therefore, there is no mass shift of $\rho_0$.

Fig.~\ref{fig-2} shows the $\rho_0$ meson
density in a magnetised medium, scaled with its density at zero
field. The shaded band
represents the uncertainty arising from neglecting $\text{Re}~ \Pi$.
This preliminary result supports the underlying assumption of our
detailed balance treatment: the magnetic field does affect the density
of neutral resonances. 
It also shows the treatment’s naivety, since the effect appears only at large field strengths and is non-monotonous.
Further work is in progress to incorporate $\pi$–$\pi$ scattering data
for a unitary, model-independent treatment of hadron
interactions~\cite{Bithika-et-Pok}.\\

{\it{Acknowledgements}}: Support by the program Excellence Initiative–Research University of the University of Wroc\l{}aw of the Ministry of Education and Science
(P.H, M.M. and B.K.), the Polish National Science Center (NCN) under
the Preludium grant 2020/37/N/ST2/00367 (M.S.) and the OPUS Grant
No.~2022/45/B/ST2/01527 (C.S., K.R. and P.M.L.), the Polish Ministry of
Science and Higher Education (K.R.), and the World Premier
International Research Center Initiative (WPI) under MEXT, Japan (C.S.)
is gratefully acknowledged.

%
%
%

\end{document}